\documentclass[journal=ancac3,manuscript=article]{achemso}

\setkeys{acs}{keywords = true}
\pdfoutput=1
\usepackage[version=3]{mhchem} 
\usepackage{amssymb}
\usepackage{amsfonts}
\usepackage{bm}
\usepackage{graphicx}
\usepackage{epstopdf}
\usepackage{color}
\usepackage{textcomp}
\usepackage{amsmath}
\usepackage{enumitem}
\definecolor{correct}{rgb}{0.85,0.3,0}

\newcommand{\GGamma}{\includegraphics{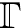}}

\title{Metasurface optical characterization using quadriwave lateral shearing interferometry}

\author{Samira Khadir}
\affiliation {Universit\'e Cote d'Azur, CNRS, CRHEA, Rue Bernard Gregory, Sophia Antipolis 06560 Valbonne, France}
\alsoaffiliation {Institut Fresnel, CNRS, Aix Marseille University, Centrale Marseille, Marseille, France}
\email{samira.khadir@chrea.cnrs.fr}
\author{Daniel Andr\'en}
\affiliation{Department of Physics, Chalmers University of Technology, 412 96 G\"oteborg, Sweden}
\author{Ruggero Verre}
\affiliation{Department of Physics, Chalmers University of Technology, 412 96 G\"oteborg, Sweden}
\author{Qinghua Song}
\affiliation {Universit\'e Cote d'Azur, CNRS, CRHEA, Rue Bernard Gregory, Sophia Antipolis 06560 Valbonne, France}
\author{Serge Monneret}
\affiliation {Institut Fresnel, CNRS, Aix Marseille University, Centrale Marseille, Marseille, France}
\author{Patrice Genevet}
\affiliation {Universit\'e Cote d'Azur, CNRS, CRHEA, Rue Bernard Gregory, Sophia Antipolis 06560 Valbonne, France}
\author{Mikael K\"all}
\affiliation{Department of Physics, Chalmers University of Technology, 412 96 G\"oteborg, Sweden}
\author{Guillaume Baffou}
\affiliation {Institut Fresnel, CNRS, Aix Marseille University, Centrale Marseille, Marseille, France}
\email{guillaume.baffou@fresnel.fr}

\abbreviations{}

\begin{document}

\newpage\begin{abstract}
An optical metasurface consists of a dense and usually non-uniform layer of scattering nanostructures behaving as a continuous and extremely thin optical component, with predefined phase and intensity transmission/reflection profiles. To date, various sorts of metasurfaces (metallic, dielectric, Huygens-like, Pancharatman-Berry, etc.) have been introduced to design ultrathin lenses, beam deflectors, holograms, or polarizing interfaces. Their actual efficiencies depend on the ability to predict their optical properties and to fabricate non-uniform assemblies of billions of nanoscale structures on macroscopic surfaces. To further help improve the design of metasurfaces, precise and versatile post-characterization techniques need to be developed. Today, most of the techniques used to characterize metasurfaces rely on light intensity measurements.  Here, we demonstrate how quadriwave lateral shearing interferometry (QLSI), a quantitative phase microscopy technique, can easily achieve full optical characterization of metasurfaces of any kind, as it can probe the local phase imparted by a metasurface with high sensitivity and spatial resolution. As a means to illustrate the versatility of this technique, we present measurements on two types of metasurfaces, namely Pancharatnam-Berry and effective-refractive-index metasurfaces, and present results on uniform metasurfaces, metalenses and deflectors.

\end{abstract}
\\


An optical metasurface consists of a planar, dense distribution of scattering nanostructures with subwavelength sizes and interdistances, behaving as a continuous and extremely thin optical component.\cite{NRM5_604,RPP81_026401,S358_eaam8100} By tailoring the distribution, composition and morphology of the nanostructures, usually called meta-atoms, the effective optical properties of the metasurfaces can be spatially adjusted to control the transmitted and/or reflected light in amplitude, phase and polarization. In particular, in order to construct efficient phase optical elements, typically lenses, one has to design meta-atoms that can cover the full $0\rightarrow2\pi$ phase-shift. This goal is impossible using metaatoms consisting of single dipolar nanoparticles, the phase of which can only vary from $0$ to $\pi$, since they behave as harmonic oscillators.\cite{OE19_21748} Thus, more complex nanostructures have to be used, usually consisting of thick structures (like pillars around 1 \textmu m in height), to favor the occurrence of retardation effects, and deviate from a dipolar response. This last decade, many kinds of metasurfaces for different applications requiring flat optical components have been reported, such as metasurfaces to control reflection and refraction \cite{S334_333, Ni2012, Sun2012}, focus light and control its polarization \cite{Niv2004, Aieta2012, NC3_1198}, produce a strong photonic spin Hall effect \cite{Yin2013, Ling2015} and project holographic images \cite{Genevet2015,NC10_2986}. 

Designing a metasurface with a specific functionality usually relies on predictions obtained from numerical simulations, including basic effective index analysis,\cite{JOSAA16_1143} Huygens-metasurface designs\cite{AOM3_813} and more sophisticated optimization procedures.\cite{N8_339,OME9_1842,NP12_659} Novel and reliable characterization techniques of metasurfaces would greatly facilitate the comparison between numerical predictions and actual measurements. Possible discrepancies and errors, originating either from the design or the imperfect fabrication of the meta-atoms, could be identified and compensated for to further improve the device characteristics. To this end, few experimental characterization techniques have been proposed, and most of them rely on partial characterization or indirect measurements such as the efficiency, the focal distance and the point-spread-function for lenses\cite{NL18_4460,SA4_eaar2114,O6_805} or of the deflection angles for phase-gradient metasurfaces.\cite{NC5_5386} In optics, the most natural method to measure phase is interferometry. Typical interferometry techniques utilize two light beams (one as a reference) recombined to interfere on a detection plane \cite{Babocky2017}. Several known drawbacks of these interferometric measurements include the problems of noise and drift, leading to poor sensitivity, and complex setups alignment. To overcome these limitations, the three-beam methods was proposed, relying on a third beam to analyse the environment and changes during the experiment.\cite{Azari2014, Ollanik2019} Traditional ellipsometry has also been used to characterize metasurfaces. However, it is only restricted to metasurfaces based on the manipulation of a phase difference between orthogonal light polarizations. \cite{Chen2013} Scanning near-field optical microscopy (s-SNOM) provides subdiffraction resolution and allows for imaging of the near-field phase response of arbitrarily complex nanoparticle arrays.\cite{Neuman2015, Bohn2015} However, s-SNOM remains time-consuming and invasive, like all near-field probe techniques. In particular, the coupling mechanisms between the near-field components and the nanoprobe tip highly depends on the tip geometry, the light polarization and the scanning mode, making reliable prediction on the actual device efficiency difficult. More recently, Bouchal et al.\cite{Bouchal2019} proposed an incoherent holographic imaging technique, but it is restricted to the characterization of Pancharatman-Berry metasurfaces and meta-atoms, or more generally orthogonal polarization components. The metasurface community is still lacking a more effective, reliable and universal characterization technique that could enable parallel wafer-level testing for device reliability and detailed characterization of the device phase profiles obtained directly after the metasurface plane.

In this article, we propose to leverage on quadriwave lateral shearing interferometry (QLSI), a quantitative phase microscopy technique based on the use of a diffraction grating, to achieve optical characterization of metasurfaces of any kind. Since QLSI measures wavefront distortion with sub-nm sensitivity and with high spatial resolution (diffraction limited), it represents a powerful tool to characterize the actual transformation applied by a metasurface on a light beam. The first part of the article introduces the working principle of QLSI and the experimental configuration. In the second part, we introduce the two families of metasurfaces that will be investigated throughout the article and that are commonly used for realizing efficient dielectric metasurfaces, namely effective-refractive-index (ERI) and Pancharatnam-Berry (PB)  metasurfaces. Then, the use of QLSI to characterize metasurfaces is exemplified first on uniform meta-atom distributions to explain the principle and demonstrate the reliability of the technique, and then on two practical cases: beam deflectors and metalenses.

\section{Quadriwave lateral shearing interferometry (QLSI)}
\subsection{QLSI working principle}

QLSI is a quantitative phase imaging technique based on the use of a wavefront analyser composed of two simple elements: a regular camera and a 2-dimensional diffraction grating, separated by millimetric distance from each other.\cite{AO39_5715} Upon illumination, the diffraction grating (usually called a modified Hartmann mask, MHM) creates an interferogram on the camera sensor that can be processed to retrieve both the intensity and the wavefront profiles $W(x,y)$, or equivalently the phase $\varphi(x,y)$,  of an incoming light beam.\cite{AO51_5698} When mounted on a microscope, the measured wavefront profile is nothing but the optical path difference (OPD) image $\delta\ell(x,y)$ created by a sample in the object plane. $\delta\ell=W$, and one usually defines
\begin{equation}
\delta\ell=\frac{\lambda_0}{2\pi}\varphi
\label{ellphi}
\end{equation}
where $\lambda_0$ is the illumination wavelength and $\varphi(x,y)$ is the phase delay experienced by a light beam crossing the sample in the object plane. The QLSI camera used in this study (Sid4 sC8 from Phasics S.A.) features an OPD sensitivity of 0.3 nm$\cdot$Hz$^{-1/2}$ corresponding to around 3 mrad$\cdot$Hz$^{-1/2}$ of phase delay in the visible range. Regarding the spatial (or lateral) resolution, it is limited by the diffraction limit associated with the numerical aperture of the objective lens, like in any optical microscopy technique. Importantly, QLSI benefits from the high sensitivity of interferometric methods but do not suffer from their usual drawbacks: it neither requires a reference beam, nor a complex alignment that might be sensitive to external perturbations. The relative positioning of the MHM with respect to the camera is done once and for all, and is not sensitive to, e.g., temperature variation, mechanical drift or air flow. Over the last decade, several applications of QLSI have been demonstrated in biology and photonics, including cell imaging,\cite{Bon2009} temperature imaging in nanoplasmonics,\cite{Baffou2012} 2D-material imaging\cite{ACSP4_3130}  and single nanoparticle optical characterization.\cite{Optica2020} This work provides another application of QLSI for nanophotonics.

\subsection{Experimental setup}

\begin{figure*}[t]
\centering
\includegraphics[scale=1]{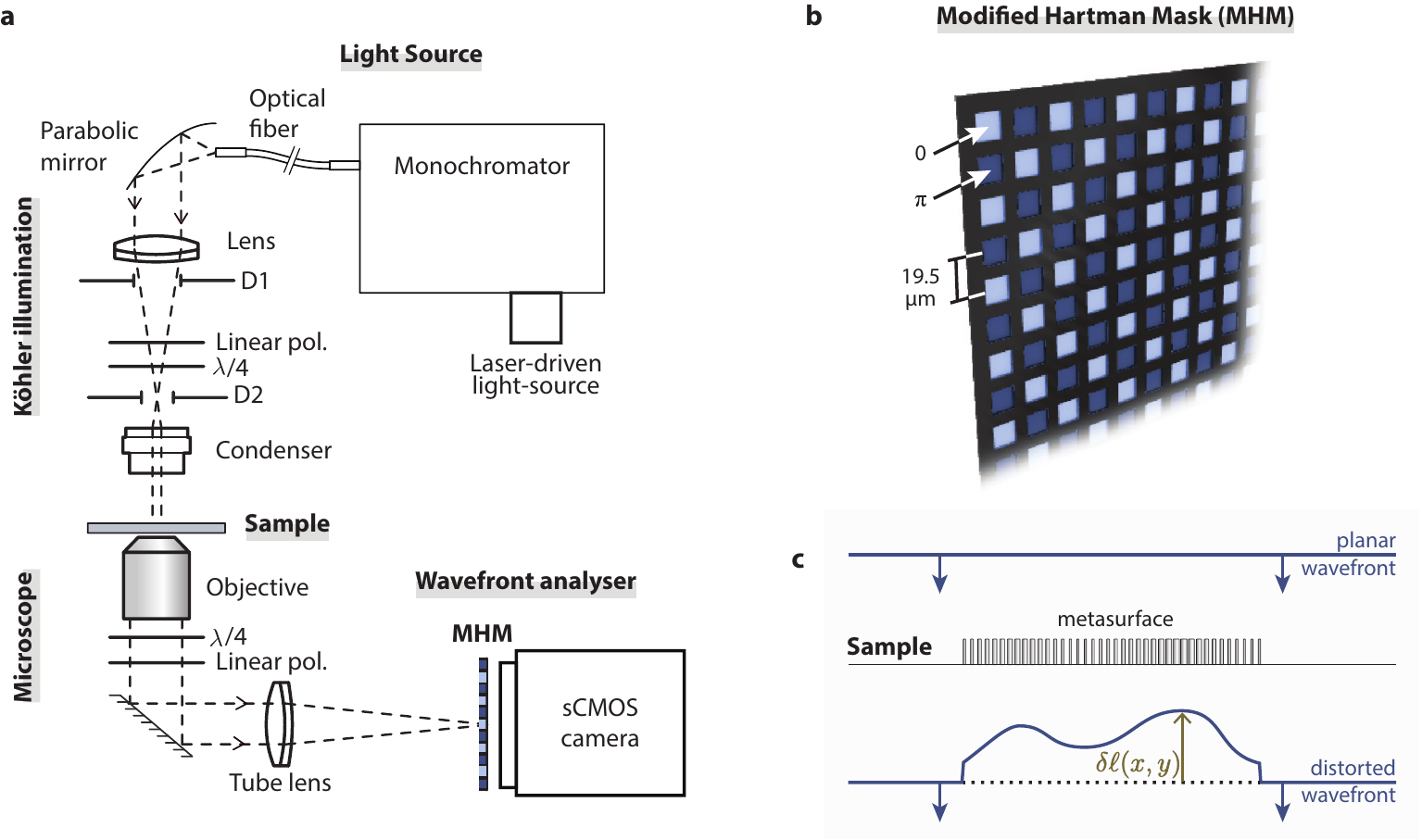}
\caption{{\bfseries Working principle of QLSI}. (a) Scheme of a QLSI microscope setup. A monochromator plasma source combined to a K\"ohler configuration illuminates the sample with a light beam controlled in wavelength, size and numerical aperture. The light passing through the metasurface sample is collected by a microscope objective lens and sent to the QLSI wavefront analyzer, composed of a sCMOS camera equipped with a modified Hartman mask (MHM). (b) Geometry of the modified Hartman Mask. (c) Schematic showing the wavefront distortion $\delta(x,y)$ experiences by a collimated light beam due to the presence of a metasurface.}
\label{setup}
\end{figure*}

The experimental configuration of the optical microscope is depicted in Fig. \ref{setup}. The illumination part consists of a plasma laser-driven-light-source (EQ-99X from Energetiq  Technologies) combined with a monochromator (Hypermonochromator, Mountain Photonics GmbH, purchased from Opton Laser International) enabling a variation of the wavelength over the visible range with a 6 nm band width and a few mW power. The use of an incoherent light source is preferable to avoid the appearance of fringes and speckles on the images. The optical fiber from the monochromator was positioned in a K\"ohler optical scheme to illuminate the sample with a sufficient degree of spatial coherence. This caution is a requirement in QLSI to achieve interferometric measurements despite the incoherent nature of the light source. The light passing through the sample is imaged by the microscope on the QLSI wavefront analyzer. Each measurement requires the acquisition of a reference image over a clear area (without any object) prior to taking an image with the object of interest within the field of view (in our case a metasurface). The reference is then subtracted from the object image to discard any imperfections of the incoming light beam. In the specific case of Pancharatnam-Berry metasurfaces characterization, a set of two optical polarizers and two quarter waveplates are added to the setup to study the device response in the standard circular cross-polarization configuration.

\subsection{Metasurfaces principles and designs}
\begin{figure*}
\centering
\includegraphics[scale=1]{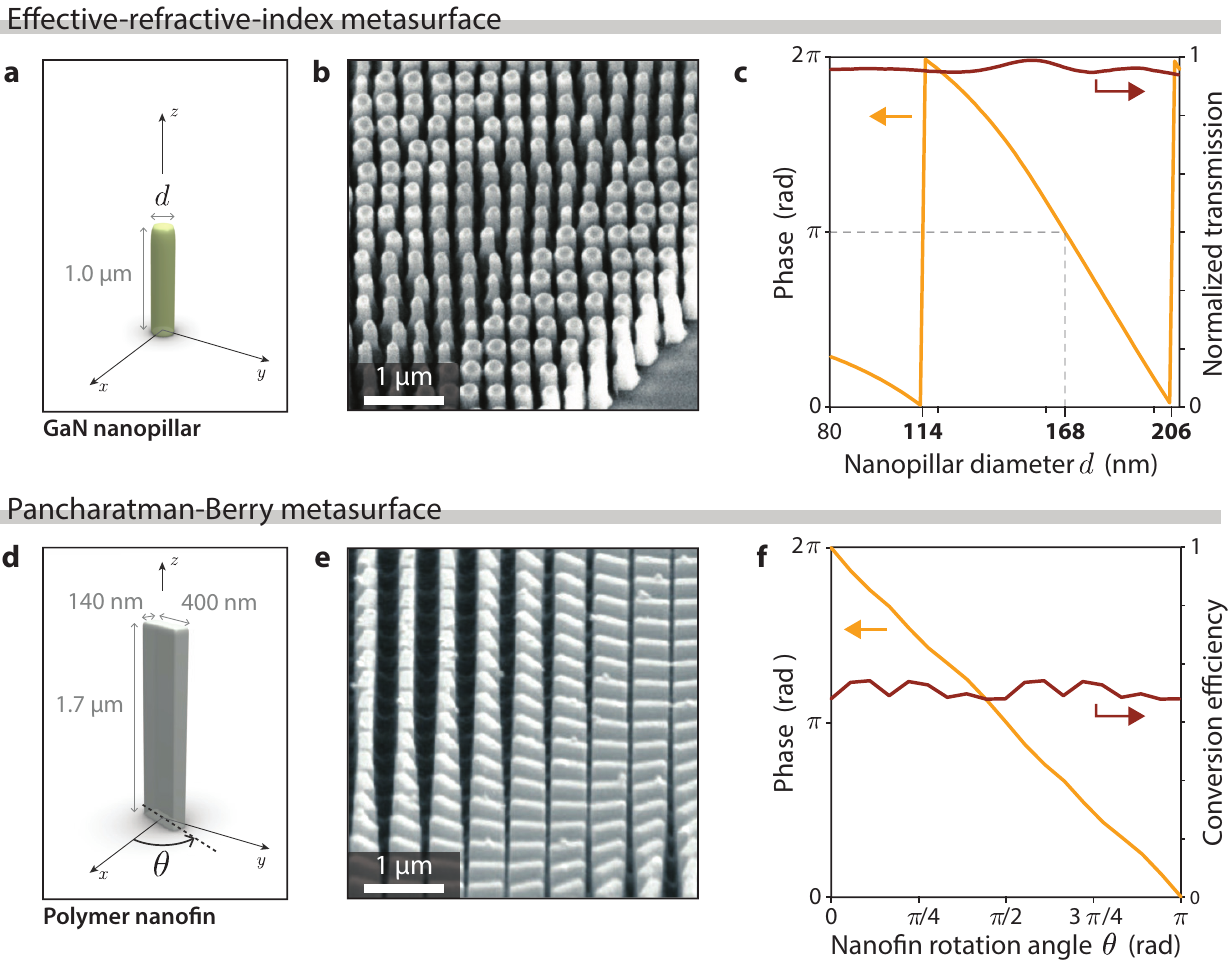}
\caption{{\bfseries Metasurface samples descriptions.} (a) Geometry of a GaN nanopillar. (b) SEM image of a ERI metasurface. (c) FDTD simulations, in periodic boundary conditions, of phase and transmittance of a GaN nanopillar array as a function of the cylinder diameter $d$, illuminated at $\lambda=600$ nm. (d) Geometry of a polymer nanofin. (e) SEM image of PB metasurface. (f) FDTD Simulations, in periodic conditions, of phase and transmittance of a nanofin array with different rotation angles illuminated at $\lambda=532$ nm.}
\label{metasurfaces}
\end{figure*}

As a means to demonstrate the versatility of QLSI for metasurfaces characterization, two opposite families of metasurfaces have been investigated, namely effective-refractive-index metasurfaces (Fig. \ref{metasurfaces}a), where the transmitted phase depends on the size of the meta-atoms, and Pancharatnam-Berry metasurfaces (Fig. \ref{metasurfaces}b), where the phase depends on the orientation of the meta-atoms.

\paragraph{\bfseries Effective-refractive-index metasurfaces.} Effective-refractive-index (ERI) metasurfaces are often composed of a dense distribution of cylindrical pillars operating as independent Fabry-Perot resonators with low quality factor (Fig. \ref{metasurfaces}a). The pillars are sufficiently tall to accommodate internal multi-longitudinal-mode propagation but remains sufficiently narrow to achieve large pillar density and prevent propagation of non-zero diffraction orders in free space or in the substrate.\cite{Genevet2017}. The resulting effective refractive index of the pillar layer, and thus the phase delay, can be adjusted by varying the pillar diameter. In this study, we used a periodic square array of nanopillars made of GaN (see Fig. \ref{metasurfaces}b and Methods part), 1 \textmu m in height and 300 nm pitch, with diameters varying from 114 to 206 nm (see Fig. \ref{metasurfaces}c).

\paragraph{\bfseries Pancharatman-Berry metasurfaces.} The Pancharatman-Berry (PB) phase is a kind of geometrical phase acquired by a light beam when passing through a birefringent medium.\cite{NR1_437} PB metasurfaces are composed of asymmetric meta-atoms usually consisting of dielectric nanofins and providing the metasurface with birefrigent properties (Fig. \ref{metasurfaces}b).\cite{S352_1190,NN13_227,NL18_4460} To achieve maximum efficiency, PB metasurfaces require the use of circularly polarized illumination, and detection of the circularly polarized light with the opposite handedness (see linear polarizers and quarter waveplates in Fig. \ref{setup}). Thus, in this cross-polarization configuration, no light is supposed to be detected, except where the sample features some birefringence, that is where PB meta-atoms are present. The action of the nanofin on a light beam can be simply modeled assuming the nanofins act as a polarizing plate associated with the following Jones matrix:
\begin{equation}
\GGamma(\beta)=\begin{pmatrix}
e^{i\beta/2} & 0\\
0 & e^{-i\beta/2}
\end{pmatrix}\label{eq:Ggamma}
\end{equation}
where $\beta$ characterizes the meta-atom anisotropy. In this case, the complex amplitude $A$ of the detected light, in the circular cross-polarized configuration, is given by (see derivation in Suppl. Info.):
\begin{equation}
A(\theta)=ie^{2i\theta}\sin(\beta/2)
\label{A_PB}
\end{equation}
where $\theta$ is the angle of the nanofin and $\sin(\beta/2)$ is called the conversion efficiency. As expected, for non-birefringent structures ($\beta=0$), no light is detected ($A=0$) due to the crossed polarization configuration in illumination and detection. The value of $\beta$ depends on the morphology, which can be optimized, at a given wavelength, to approach as much as possible $\sin(\beta/2)=1$, i.e., $\beta=\pi$, which means that an ideal PB meta-atoms should act as a half-wave plate ($\GGamma(\pi)$). In this study, we used nanofins made of patternable polymer with a pitch of 500 nm (see Fig. \ref{metasurfaces}d-f and Methods part).

The transmission of light with the opposite handedness is due to the presence of birefringent structures and Eq.\eqref{A_PB} indicates that the phase of this transmitted light is dictated by the orientation of the nanofins. Setting the nanofin angle to $\theta$ results in a change of the local geometric phase of $2\theta$, no matter the composition or the morphology of the nanofin, providing it has some anisotropy.\\



The phase of ERI and PB meta-atoms are adjusted by varying respectively their size and their angle. In most nanofabrication techniques, a diameter is more difficult to accurately control compared to an angle.  For instance, different e-beam doses are markedly affecting the sizes of lithographically fabricated objects, but not their orientations. For this reason, ERI metasurfaces are particularly sensitive to nanofabrication inaccuracies. Any inaccuracy in PB meta-atom size only affects the conversion efficiency ($\sin(\beta/2)$ in Eq. \eqref{A_PB}), i.e. the transmitted intensity, not the phase shift. Here lies the benefit of PB metasurfaces. PB metasurface are also dispersive-less (i.e., achromatic) since the phase of $A(\theta)$ (Eq. \eqref{A_PB}) is wavelength-independent, unlike ERI metasurfaces. However, PB metasurfaces are not outperforming ERI metasurfaces in any aspect. The main limitation of PB metasurfaces is the necessity to work with circularly polarized light to achieve maximum efficiency. In the following, we illustrate and discuss the use of QLSI for these two families of metasurfaces.

\section{Results and Discussion}

\subsection{Characterization of uniform metasurfaces}

\begin{figure*}[h]
\centering
\includegraphics[scale=1]{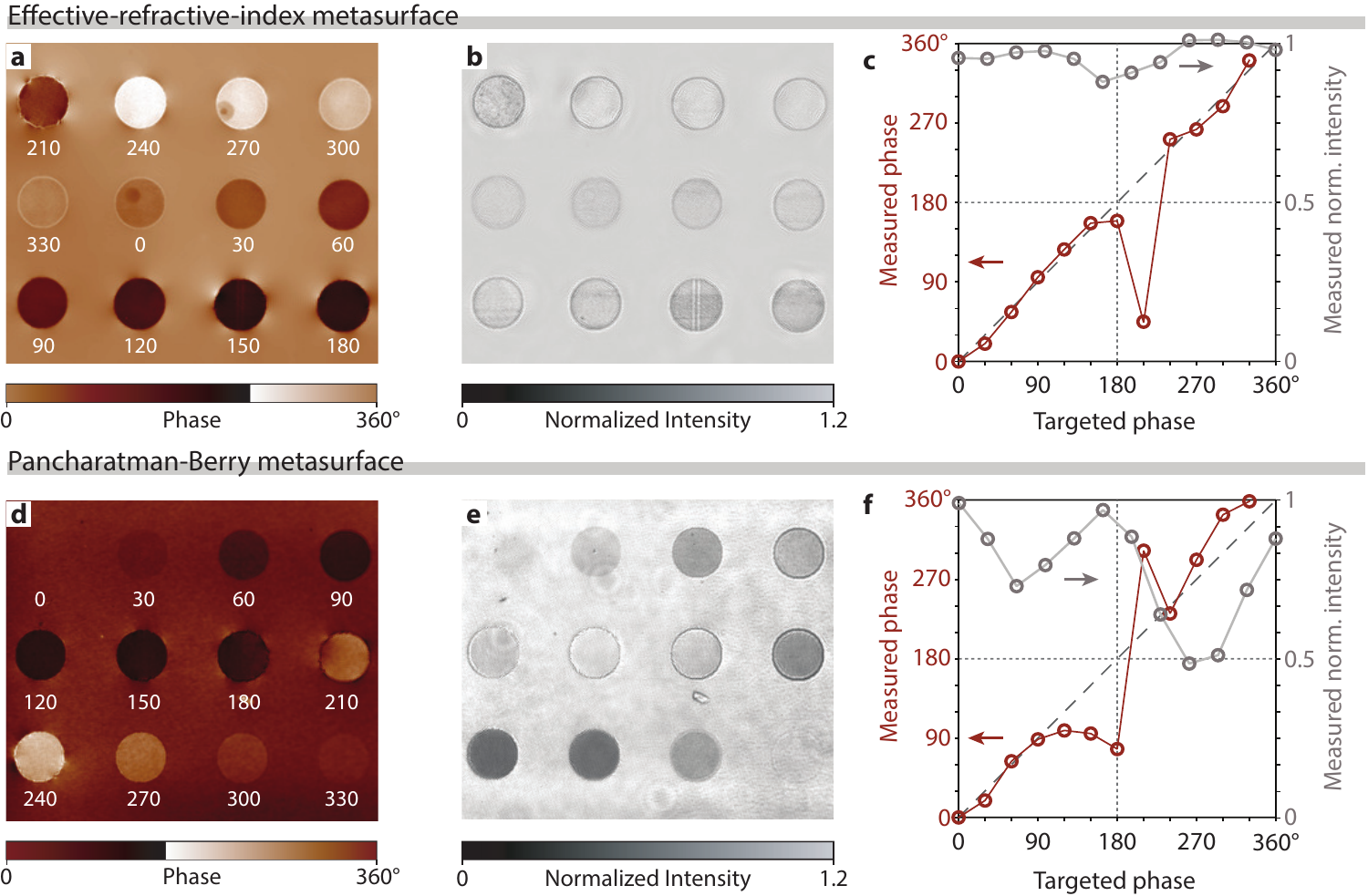}
\caption{{\bfseries QLSI characterization of uniform metasurfaces}. (a) Phase and (b) intensity images of 12 circular arrays (30 \textmu m in diameter) of different GaN nanopillars, the sizes of which increases gradually from left to right and then from top to bottom. The targeted phase values are indicated on the phase image for each area, varying by steps of $30^\circ$. (c) measurements of the phase and normalized intensity of the circular metasurfaces shown in images (a) and (b). (d) Phase and (e) intensity images of 12 circular arrays (30 \textmu m in diameter) of different polymer nanofins, the orientation of which varying gradually by steps of $15^\circ$ to make the phase vary by steps of $30^\circ$. The targeted phase values are indicated on the phase image for each area. (f) measurements of the phase and normalized intensity of the circular metasurfaces shown in images (d) and (e).}
\label{UniformArrays}
\end{figure*}

Figure \ref{UniformArrays} presents QLSI images of \emph{uniform} metasurfaces. In both case studies (ERI and PB), 12 circular "phase-piston" metasurfaces have been imaged within a single field of view. Each metasurface corresponds to a given phase shift, varying gradually by steps of $30^\circ$ to span the entire $2\pi$ range.

The case of ERI metasurfaces corresponds to Figs. \ref{UniformArrays}a--c. The phase shifts are uniform over each disc and the values are reported in Fig. \ref{UniformArrays}c. The agreement with the targeted phase is very good, except for phase shifts close to $\pi$. This issue is expected: QLSI primarily measures phase gradient images over $x$ and $y$ directions. The phase image is subsequently reconstructed numerically from these two gradient images. When a phase jump close to $\pi$ occurs between neighboring pixels in a gradient image, the phase reconstruction algorithm cannot really know if the phase is actually decreasing or increasing, and may assign a phase gradient of $\pm\pi$, the sign of which being highly noise dependent. This random sign attribution can create some rendering issues, such as discontinuities and singularities in the phase image, like in the top-left area of Fig. \ref{UniformArrays}a. However, this issue is not supposed to be a problem for most metasurfaces characterization as metasurfaces rarely exhibit phase steps of $\pi$, and generally rather consist of regular and smooth variations of phase over space (see the metalens and metagradient characterization hereinafter). Fixing this issue would require the implementation of refined phase reconstruction algorithms. QLSI also enables the reconstruction of the intensity image from the interferogram image. The normalized intensity image is displayed in Fig. \ref{UniformArrays}b and measurements reported in Fig. \ref{UniformArrays}c, which has to be compared with Fig. \ref{metasurfaces}c. The transmission of the metasurfaces is close to theoretical prediction, i.e., quite constant upon varying the nanopillar diameter.

The case study of PB uniform metasurfaces is presented in Figs. \ref{UniformArrays}d--f. The sample contained 12 uniform and circular metasurfaces, in which the nanofin angle varied from one metasurface to another by increments of $15^\circ$ to generate phase increments of $30^\circ$ and cover the full $2\pi$ phase shift. The circular areas observed in the images were surrounded by a uniform sea of fins oriented as in the $0^\circ$-area, as a means to get non-zero light intensity outside the circular areas and also to have a nearby blank area with some transmitted light to acquire the reference image. The absence of a layer of PB meta-atoms around the metalens would have resulted in a zero light intensity background, which is likely to create large noise on the image and even rendering issues upon reconstruction of the phase image. The phase values are reported in Fig. \ref{UniformArrays}f, with a good agreement with theory, except, again, when the phase is close to $\pi$, for the reason mentioned above. The normalized intensity image is displayed in Fig. \ref{UniformArrays}e and measurements reported in Fig. \ref{UniformArrays}f, which have to be compared with Fig. \ref{metasurfaces}f. The transmission of the metasurfaces is not as uniform as expected upon varying the nanofin angle. This discrepancy can result from near-field interaction between neighboring fins, or possibly from some misalignment of the illumination and detection polarizing plates.

\subsection{Characterization of beam deflectors}
We now focus the study on a textbook case that is a beam deflecting, or beam steering metasurface.\cite{S334_333,NC5_5386,S345_298,SR3_2155} Again, we studied deflectors made of both ERI and PB metasurfaces. The devices have been designed to deflect light on the air side at a fixed angle of $\theta=5^\circ$. Considering $(Oy)$ as the direction of deflection, the theoretical OPD reads
\begin{equation}
\delta\ell(y)=y\,n\,\sin(\theta)
\label{deflector}
\end{equation}
where $n$ is the refractive index of the medium ($n=1$ in our case).

\begin{figure}[t]
\centering
\includegraphics[scale=0.7]{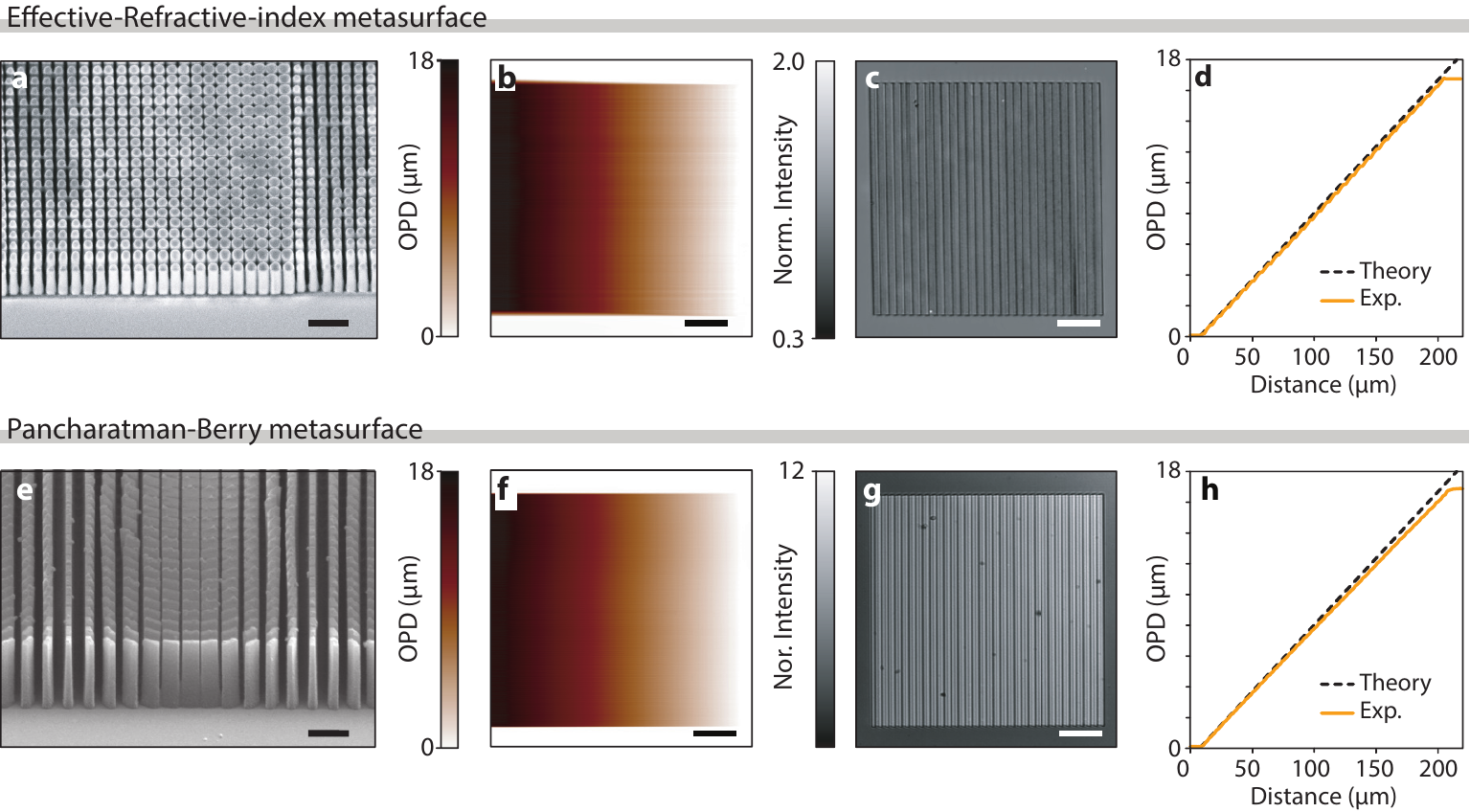} 
\caption{{\bfseries QLSI characterization of beam deflectors}. (a) SEM image of a ERI deflecting metasurface (scale bar: 1 \textmu m). (b) OPD image of the ERI beam deflector. (c) Associated normalized intensity image. (d) Averaged horizontal profile of (b) (solid line), compared with theory (dashed line). (e-h) Same as (a-d) for a PB beam deflecting metasurface.}
\label{deflectors}
\end{figure}

Experimental results are presented in Fig. \ref{deflectors}. SEM images (Figs. \ref{deflectors} a, e) detail gradual distribution of meta-atoms at the nanoscale. OPD images (Figs. \ref{deflectors}b,f) display the expected linear gradient of phase, in very good agreement with  theory (Figs. \ref{deflectors}d,h). Intensity images (Figs. \ref{deflectors}c,g)  display some non-uniformities, but which do not noticeably affect the phase gradient.

\subsection{Characterization of metalenses}

We focus now on the most common application of a metasurface, which is the metalens. We studied equivalent ERI and PB metalenses, designed with a fixed radius of $R=100$ $\mu $m and a focal length of $f=500$  \textmu m (NA$\sim$0.2). The theoretical OPD profile of such lenses is given by: 

\begin{equation}
\delta\ell(r)=\mp\sqrt{(r^2+f^2)} + C
\label{metalensEq}
\end{equation}
where $r$ is a radial coordinate and $C$ a constant. The signs $-$ or $+$ correspond to convergent or divergent metalenses, respectively. 

\begin{figure*}
\centering
\includegraphics[scale=1]{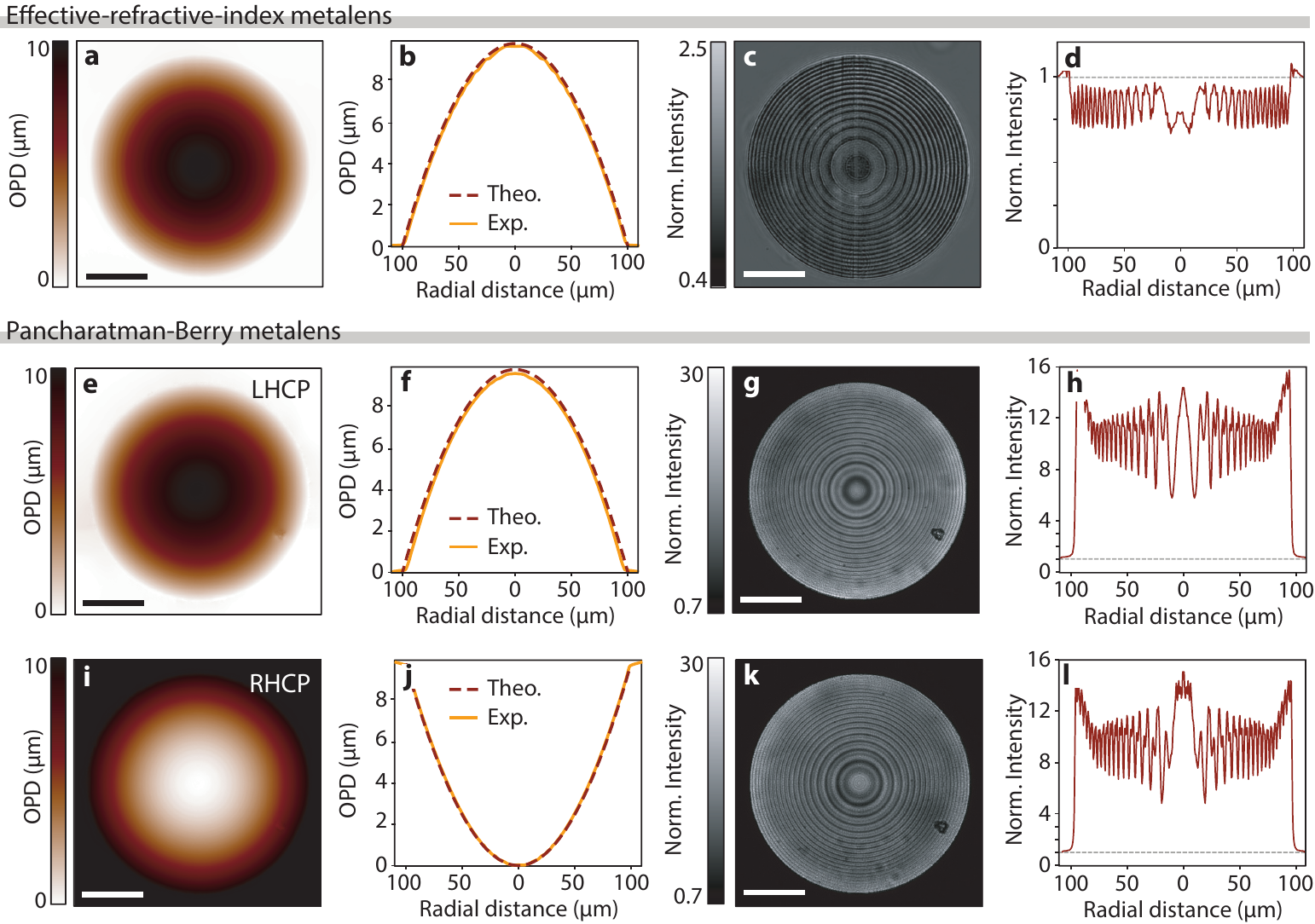}
\caption{{\bfseries QLSI characterization of metalenses}. (a) OPD images of the ERI metalens recorded by QLSI, along with its (b) radially averaged profile. (c) Normalized intensity image recorded by QLSI, associated to image (a), along with its (d) radially averaged profile. (e-h) Same as (a-d) for the PB metalens illuminated by a left-handed circular polarization (LHCP) beam. (i-l) Same as (e-h) the PB metalens illuminated by a RHCP beam. Scale bars for QLSI images: 50 $\mu$m.}
\label{metalenses}
\end{figure*}

Figures \ref{metalenses}a--d present results on the characterization of a ERI metalens using QLSI. The radially-averaged OPD profile (Fig.\ref{metalenses}b) shows a perfect agreement with theory (see Eq. \eqref{metalensEq}). The intensity profile (Fig.\ref{metalenses}d) is not perfectly uniform, exhibiting spatial variations in agreement with Fig. \ref{UniformArrays}c.

Figures \ref{metalenses}e--l present results on the characterization of a PB metalens. Again, phase measurements (Fig. \ref{metalenses}e) perfectly match theoretical predictions (Fig. \ref{metalenses}f). Interestingly, for PB metalenses, reversing the handedness of both the illumination and detection circular polarizations leads to a metalens with the opposite focal length $-f$.\cite{NC3_1198} This effect is indeed observed and characterized in Fig. \ref{metalenses}i, which reproduces the characteristics of a diverging lens in excellent agreement with theoretical predictions (Fig. \ref{metalenses}j). Just like with the ERI metalens, unexpected features are observed in the transmission profiles (Fig.\ref{metalenses}h,l): While theory predicts a transmission that is independent of the fin orientation (see Eq. \eqref{A_PB}), variations are observed throughout the metalens in agreement with Fig. \ref{UniformArrays}f. We evidence here that this non-uniform transmittance does not affect the quality of the phase profile.

\begin{figure}[!ht]
\centering
\includegraphics[scale=1]{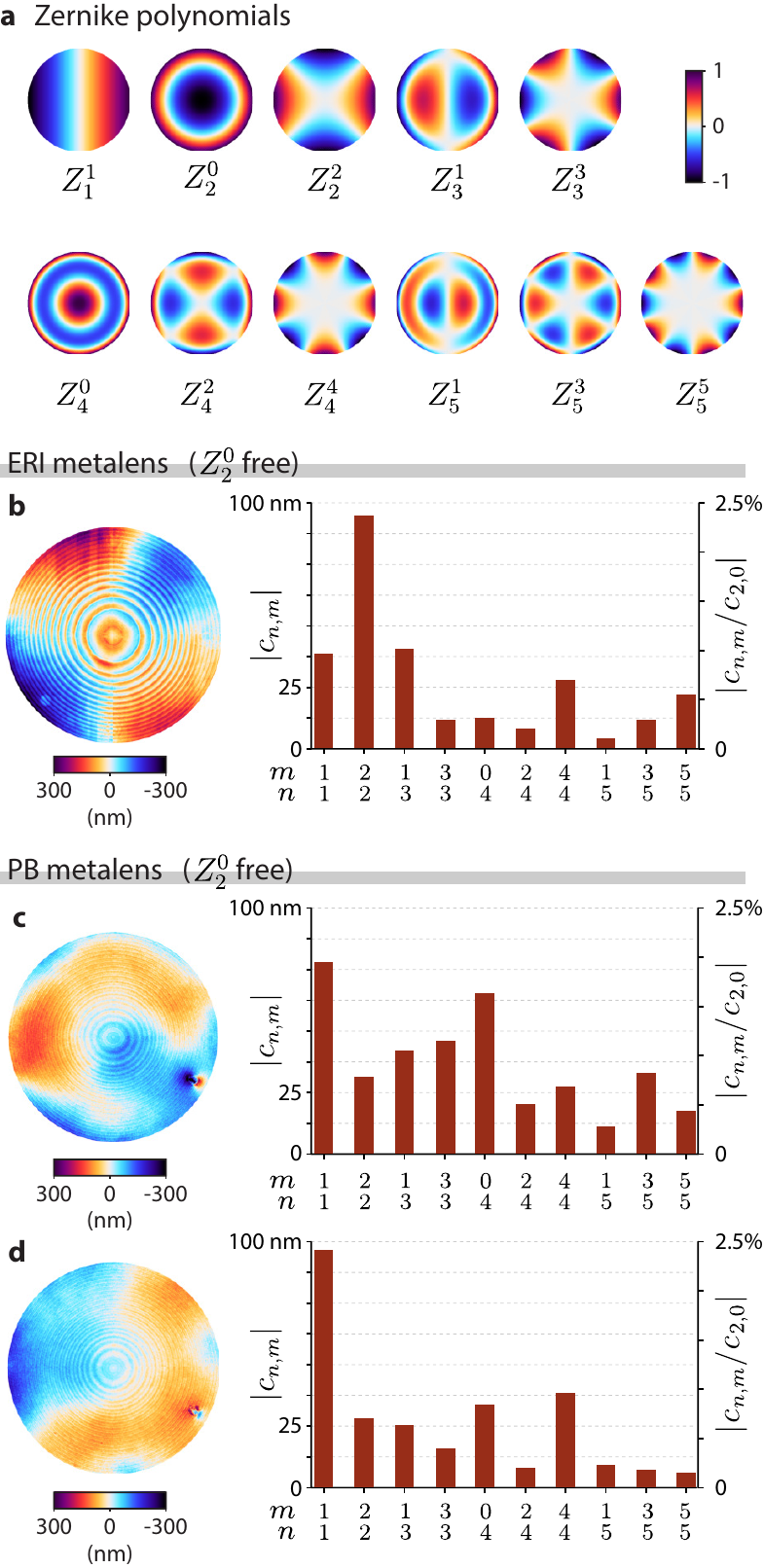}
\caption{{\bfseries QLSI and Zernike analysis of metalenses.}(a) Numerical simulations of the real parts of Zernike polynomials up to $n=5$. (b) left: optical thickness of an ERI metalens, from which the Zernike mode $Z_2^0$ was subtracted, highlighted the lens aberrations and imperfections. (right): measurements of the Zernike moments up to $n=5$. (c) left: optical thickness of a PB metalens, from which the Zernike mode $Z_2^2$ was subtracted. (right): measurements of the Zernike moments up to $n=5$. (d) Same as (c) for the opposite circularly-polarized illumination handedness.}
\label{Zernike}
\end{figure}

Measurements of Zernike moments is common to characterize the aberrations of an optical system with circular pupils.\cite{JMO58_545,NP13_649} We explain hereinafter how QLSI can efficiently measure Zernike moments of metasurfaces. The Zernike polynomials $Z_n^m$ are defined over the unit disk $r\in[0,1]$, $\phi\in[0,2\pi]$ as 
\begin{equation}
    Z_n^m(r,\theta)=R_n^m(r)e^{im\phi}
\end{equation}
$n$ and $m$ are integers, following the restrictions $n\ge0$, $0\le m\le n$, and $m$ of same parity as $n$. $R_n^m(r)$ are polynomials with real coefficients, called the radial Zernike polynomials. For instance $R_1^1(r)=r$, $R_2^0(r)=2r^2 - 1$ and $R_2^2(r)=r^2$. Any smooth, real function $F$ defined over a disc can be written as a linear combination of Zernike polynomials $(Z_n^m)_{n,m\in\mathbb{N}}$, which form an orthonormal image basis.
\begin{equation}
F=\frac{1}{2}\sum_{n,m}c_{n,m}Z_n^m + c_{n,m}^\star Z_n^{-m}
\end{equation}
$c_{n,m}$ are complex coefficients called the complex Zernike moments of the image. The first elements of the Zernike basis, corresponding to the lowest spatial frequencies of the image, are represented in Fig. \ref{Zernike}a. Calculation of the moments of an image such as the phase image represented in Figs. \ref{metalenses}a,e,i requires to precisely define the center and the radius of the disc over which the moments have to be calculated. Then, the Zernike moments can be retrieved using
\begin{equation}
    c_{n,m}=\frac{2(n+1)}{\pi\epsilon_m}\int_0^{2\pi}\int_0^1F(r,\phi)Z_n^m(r,\phi)r\mathrm{d}r\mathrm{d}\phi
\end{equation}
where $\epsilon_m=2$ for even $m$ values and $\epsilon_m=1$ for odd $m$ values: $\epsilon_m=2-\mathrm{modulo}(m,2)$.

$Z_1^1(r,\phi)=re^{i\phi}$ represents a tilt over the direction $\phi$. This moment is particular and requires some comments. First, the moment strongly depends on the center of the disc on which the moment is calculated. An imprecise determination of this position creates an artificial $c_{1,1}$ moment. Second, this moment is not really important for the characterization of the metalens as it affects neither the point spread function (PSF), nor the efficiency of the metalens. The effect will only be a shift of the PSF. For these reasons, the tilt moment $c_{1,1}$ is of lesser interest compared to the other moments.

The second polynomial that requires attention, especially for a metalens, is $Z_2^0(r,\phi)=2r^2-1$. When applied to an image obtained with an imaging system, $Z_2^0$ normally corresponds to the aberration called \emph{defocus}. But when characterizing a lens, it is nothing but the moment corresponding to the phase profile of the lens itself, at least at the first order. Besides, the focal length of the metalens is linked to the $Z_2^0$ moment by the expression (see derivation in Suppl. Info.):
\begin{equation}
f=-\frac{R_\mathrm{Z}^2}{4c_{2,0}}
\label{fZa}
\end{equation}
$R_\mathrm{Z}^2$ is the radius of the disc over which the Zernike moment $c_{2,0}$ is calculated. Ideally, it should be the radius of the metalens so that the full surface of the metalens is characterized. Note that the phase profile of the lens is hyperbolic (see Eq.\eqref{metalensEq}) while $Z_2^0$ is parabolic. Thus, the fit is not perfect unless one has $R\ll f$, which is usually the case. In that case, one can assume a first order approximation of Eq.\eqref{metalensEq} giving a parabolic profile. Eq. \eqref{fZa} was derived using this approximation (see Suppl. Info.).

The following Zernike moments $c_{2,2}$, $c_{3,1}$, $c_{3,3}$, \dots\ are estimations of deviations from the perfect parabolic profile of $Z_2^0$. These deviations can come from imperfections of the metalens. It can also come from imperfections of the illumination in case polarization matters (like with PB metasurfaces). $c_{n,0}$ can also come from the deviation of the hyperbolic profile from a parabolic profile, especially when the radius of the metalens is not significantly smaller than the focal length, i.e., for metalenses with large numerical apertures.

Figure \ref{Zernike}b(left) displays the optical thickness of a ERI metalens measured by QLSI, from which the measured $Z_2^0$ polynomial has been removed, as a means to highlight the imperfections. A dominant $Z_2^2$ aberration is observed, corresponding to astigmatism. This dominant aberration is also evidenced in Fig. \ref{Zernike}b(right) reporting all the Zernike moments. All the other moments are very small, smaller than 1\% of the $(2,0)$ moment.  The moments' moduli are also plotted in nanometers, to give an idea of the expected signal, as compared with the camera sensitivity. Since QLSI sensitivity is around 1 nm, all the Zernike moments reported in Fig. \ref{Zernike}b(right) are well above the sensitivity of the technique, albeit very small, and do not correspond to measurement noise. All these aberrations come from the metalens only, and are not supposed to come from any phase imperfection of the incoming light. Imperfections of the incoming light are unavoidable, but one gets rid of them in QLSI by subtracting a reference image.

Figures \ref{Zernike}c,d address the case of a PB metalens, used as a convergent (Fig. \ref{Zernike}c) or divergent (Fig. \ref{Zernike}d) lens by rotating the circular polarizations. The difference with ERI metasurfaces is here that the Zernike moments can also depend on the light beam quality: any non-uniform deviation from a perfect circular polarization over the field of view may result in additional phase non-uniformity. The acquisition of a reference image may not compensate for this problem. Thus, the measured Zernike moments characterize both the light beam and the metalens, and cannot be only attributed to metalens aberrations. This rule applies any time the polarization (linear or circular) matters for the characterization of metalenses, but can be more problematic for PB metasurfaces because perfectly circular polarizations are more complicated to achieve compared with linear polarizations. Any QLSI experiment on PB metasurfaces investigating high-order Zernike coefficients has to carefully set the circular polarizations in illumination and detection in order to faithfully probe the metalens quality and not something else.\\

\begin{figure}[t]
\centering
\includegraphics[scale=1]{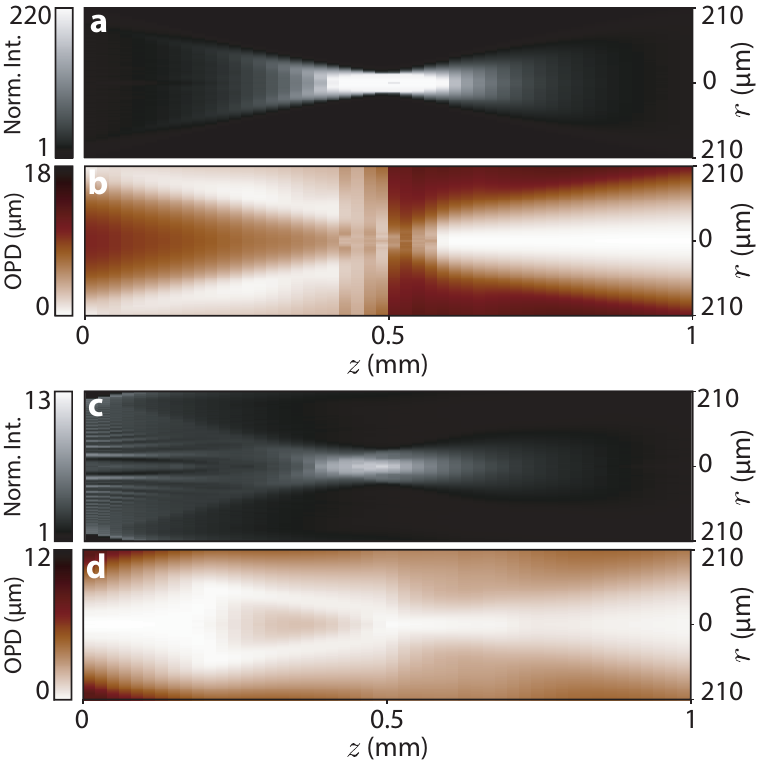}
\caption{{\bfseries QLSI measurements of light propagation after a PB metalens}. (a,b) LHCP/RHCP and (b,c) RHCP/LHCP configurations. (a) and (c) are intensity measurements, (b) and (d) are OPD measurements.}
\label{metalensfocus}
\end{figure}

QLSI imaging is not limited to the focus plane. The technique can image the phase profile anywhere in space. In particular, QLSI can also characterize the light propagation in phase and intensity after crossing the metasurface, as a means to better characterize the action of a metasurface. For this purpose, we translated the objective lens along its optical axis $(Oz)$ and acquired several images as a function of its position along $(Oz)$. An equivalent option could have been to translate the sample along $(Oz)$. The intensity image presented in Fig.\ref{metalensfocus}a confirms that the light beam exiting the metalens is focused at a distance that corresponds to the focal length ($f=500$ \textmu m). The OPD distribution  (Fig.\ref{metalensfocus}b) exhibits an inversion of the contrast at the focal point corresponding to a phase shift of $\pi$. This phenomenon is known as the Gouy phase shift.\cite{Feng2001} The tiny instabilities of the OPD image close to the focus spot, observed in Fig.\ref{metalensfocus}b, result from the lack of spatial coherence. At the focus, the metalens images the source, which is in our experiment the exit of the optical fiber of the monochromator that is not spatially coherent. QLSI performs less well in that condition. Figs. \ref{metalensfocus}c,d) report measurements using the reverse circular polarization configuration, where the metalens acts as a divergent lens. Regarding the light intensity (Fig. \ref{metalensfocus}c), it is vanishing upon propagation due to the divergence of the beam, with a residual, weak converging beam (1 order of magnitude lower than for the convergent lens) due to some imperfection of the illumination leading to a leakage of co-polarized light. Regarding the phase profile (Fig.\ref{metalensfocus}d, the region 0-100 nm away from the metasurface clearly exhibit a diverging behavior, as the phase profile is convex and spreading out.


\section{Conclusion}
In summary, we introduce here the use of quadriwave lateral shearing interferometry (QLSI) as a tool to fully characterize the optical properties of metasurfaces, in a simple, versatile and accurate way. We show how to apply the technique for both effective-refractive-index and Pancharatnam-Berry phase metasurfaces. Since typical optical thicknesses of metasurfaces are on the order of a few microns, while the sensitivity of QLSI is around 1 nm, one can achieve very high precision on metasurfaces characterization. 
Noteworthily, so far, quantitative phase imaging techniques have been almost only developed and used for cell imaging in biology. We believe quantitative phase imaging is on the verge to become a tool of predilection in nanophotonics, especially thanks to the unequaled high spatial resolution (diffraction limit), high sensitivity (down to 1 nm) and quantification capabilities of QLSI. All these benefits are important requirements for accurate characterization of nanophotonics components. Furthermore, the compactness and the simplicity of operation of QLSI makes it compatible with most commercial microscope and suited for with mass market production, to perform optical testing and validation of optical components during various phases of manufacturing in production lines.

\section{Methods}
\paragraph{\bfseries ERI metasurface fabrication.} Effective-refractive-index metasurfaces were made of square periodic arrays of GaN pillars, 1 \textmu m in height, 300 nm pitch. They have been revealed by patterning a 1 \textmu m thick GaN layer grown on a double-side polished c-plan sapphire substrate using a Molecular Beam Epitaxy (MBE) RIBER system. Conventional electron beam lithography (EBL) was used to expose a double layer of $\sim$200 nm PMMA resist (495A4) spin-coated on the GaN thin-film and then baked on a hot plate at 125$^\circ$C. E-beam resist exposure was then performed at 20 keV (using a Raith ElphyPlus, Zeiss Supra 40), followed by PMMA development using a 3:1 IPA:MIBK solution. After development, a 50-nm layer of Ni was deposited using e-beam evaporation, to perform a metallic film liftoff by immersing the sample into acetone solution for 2 h. The resulting Ni pattern was utilized as a hard mask during the reactive ion etching (RIE, Oxford system with a plasma composed of \ce{Cl2}, \ce{CH4}, \ce{Ar} gases, with flows of 13, 2 and 2 sccm respectively) to transfer the pattern in the GaN layer. Finally, the Ni hard mask on the top of GaN nanopillars was removed using chemical etching with a 1:2 HCl:\ce{HNO3} solution. ERI metasurfaces have been designed to be used at a wavelength of $600$ nm.

\paragraph{\bfseries PB metasurface fabrication.} We fabricated the PB metasurfaces using a novel approach based on the use of a patternable polymer (ma-N 2410, Micro Resist Technology GmbH, Germany) as the metasurface building material itself. The fabrication process is described in detail in Ref. \cite{ACSP7_885}. This fabrication technique allows for a facile route to making large-scale, high efficiency gradient metasurfaces. The constituent meta-atoms are freestanding nanofins with the dimension $400\times140\times1700$ nm (Fig. \ref{metasurfaces}), which when placed in a square lattice with a 500 nm period, present a polarization conversion efficiency of $\sin(\beta/2)\approx50\%$ at the operating wavelength of 530 nm. Using this type of metasurface building block, a hyperbolic lens with a focal length of 500 \textmu m as well as a linear phase gradient metasurface were built. In the case of uniform metasurfaces and deflectors (but not in the case of metalenses), the metasurfaces were surrounded by a uniform sea of fins, as a means to have light transmission also around the metasurfaces. All PB metasurfaces have been designed to be used at a wavelength of $532$ nm.

\begin{acknowledgement}
S. K., Q. S.and P.G. acknowledge financial support from the European Research Council (ERC) under the European Union's Horizon 2020 research and innovation programme (grant agreement FLATLIGHT no. 639109 and grant agreement i-LiDAR no. 874986). D.A., R.V. and M.K. would like to thank the Knut and Alice Wallenberg Foundation, the Swedish Foundation for Strategic Research and the Excellence Initiative Nano at Chalmers University of Technology. This work was partially performed at Myfab Chalmers. The authors thank B. Wattellier and R. Laberdesque (from Phasics company) for helpful discussions in particular about the use of the phase reconstruction software.

\end{acknowledgement}


\begin{suppinfo}
\end{suppinfo}

\bibliography{metasurfaces}

\begin{tocentry}

\end{tocentry}

\end{document}